\documentclass[12pt,journal,draftclsnofoot,letterpaper,oneside,onecolumn]{IEEEtran}

\usepackage{mathtools}
\usepackage{amsfonts}
\usepackage{mathrsfs}
\usepackage{cite}
\usepackage{graphicx}
\usepackage{amsmath}
\usepackage{amssymb}
\usepackage{stfloats}
\usepackage{subeqnarray}
\usepackage{cases}
\usepackage{indentfirst}
\usepackage{bm}
\usepackage{setspace}

\begin{document}

\newtheorem{condition}{Condition}
\newtheorem{assumption}{Assumption}
\newtheorem{colloary}{Colloary}
\newtheorem{theorem}{\bf Theorem}
\newtheorem{property}{\bf Property}
\newtheorem{proposition}{Proposition}
\newtheorem{lemma}{\bf Lemma}
\newtheorem{example}{Example}
\newtheorem{notation}{Notation}
\newtheorem{definition}{\bf Definition}
\newtheorem{remark}{Remark}

\title{\Huge{Coalitional Graph Games for Popular Content Distribution in Cognitive Radio VANETs}}

\author{
\IEEEauthorblockN{\normalsize{Tianyu Wang}\IEEEauthorrefmark{1}, \normalsize{Lingyang Song}\IEEEauthorrefmark{1}, and \normalsize{Zhu
Han}\IEEEauthorrefmark{2}\\}
\IEEEauthorblockA{\IEEEauthorrefmark{1}\normalsize{State Key Laboratory of Advanced Optical Communication Systems and Networks, Peking University, Beijing, China.} \\
\IEEEauthorrefmark{2}\normalsize{Electrical and Computer Engineering Department, University of Houston, Houston, USA.} }
}

\maketitle

\begin{abstract}

Popular content distribution is one of the key services provided by vehicular ad hoc networks (VANETs), in which a popular file is broadcasted by roadside units (RSUs) to the on-board units (OBUs) driving through a particular area. Due to fast speed and deep fading, some file packets might be lost during the vehicle-to-roadside broadcasting stage. In this paper, we propose a peer-to-peer (P2P) approach to allow the OBUs to exchange data and complement the missing packets. Specifically, we introduce a coalitional graph game to model the cooperation among OBUs and propose a coalition formation algorithm to implement the P2P approach. Moreover, cognitive radio is utilized for vehicle-to-vehicle transmissions so that the P2P approach does not require additional bandwidth. Simulation results show that the proposed approach performs better in various conditions, relative to the non-cooperative approach, in which the OBUs share no information and simply response to any data request from other OBUs.

\end{abstract}

\newpage

\section{Introduction}

\IEEEPARstart{V}ehicular ad hoc networks (VANETs) have been envisioned to provide increased convenience and efficiency to drivers, with numerous applications ranging from traffic safety, traffic efficiency to infotainment \cite{OW-2009,HL-2008}. One particular type of downloading services has attracted a lot of attentions for its applications in both safety-related and commercial areas. That is, the distribution of popular multimedia contents to on-board units (OBUs) inside a geographical area of interest (AoI) by roadside units (RSUs), which is referred to as popular content distribution (PCD) in \cite{LYL-2011}. Examples of PCD may include: a local hotel periodically broadcasts multimedia advertisements to the vehicles entering the city on suburban highway; a local travel company advertising the current activities in scenic areas to the passing vehicles; and a traffic authority delivers real-time traffic information ahead, or disseminates an update version of local GPS map \cite{LYL-2011}.

On the Internet, the downloading services of large files (e.g. high-definition movies) often adopt peer-to-peer (P2P) protocols, such as BitTorrent and eDonkey2000 \cite{LCPSL-2005}. Those P2P systems go beyond client-server systems by introducing symmetry ideas (a client may also be a server) and enjoy high performances in data rate, delay, scalability, and robustness. For PCD in VANETs, it is natural to adopt P2P ideas for improving the network performance. Actually, since it takes usually less than $1$ minute for moving vehicles to drive through the coverage of an RSU, the OBUs may fail to download the large popular file (e.g. an advertisement video may be as large as $100$ MB) directly from RSUs within the limited time for vehicle-to-roadside (V2R) transmissions. For completely downloading the popular file, it is not only optional, but also essential for the OBUs to build a self-organizing decentralized P2P network, in which popular packets are exchanged among OBUs through vehicle-to-vehicle (V2V) channels. However, those P2P techniques on the Internet should be carefully inspected before applying in the proposed PCD problem for the following reasons:
\begin{enumerate}
    \item The wireless links in VANETs are unreliable due to both deep fading and co-channel interference, compared to the wired links on the Internet.
    \item The network topology of a VANET is unpredictable and ever-changing, due to the high mobility of OBUs, compared to the static topology of the Internet.
\end{enumerate}
Hence, P2P protocols in VANETs are no longer application level protocols based on reliable transmissions, but cross-layer protocols that jointly considers content request, peer location, channel capacity, and potential interference.

In \cite{NDPGS-2005}, the author first studied cooperative downloading services in VANETs, in which they proposed SPAWN, a pull-based, P2P content downloading protocol that extends BitTorrent. However, the peer and content selection mechanisms have high overhead and are not scalable, especially when most of the vehicles are interested in downloading popular contents. In \cite{LPYPG-2006, LYL-2011}, network coding (NC) methods were proposed, in which packets are mixed together by coding at every intermediate node and the broadcast nature of wireless medium is exploited, so that the usefulness of each coded packet is increased. In \cite{ZZJ-2009}, the author focused on the link layer, and proposed VC-MAC, a cooperative medium access control (MAC) protocol for gateway downloading scenarios in vehicular networks, to maximize the ``broadcast throughput".

In this paper, we propose a P2P approach to address the PCD problem in VANETs, in which the OBUs are allowed to exchange data and complement their missing packets. Among all the possible P2P solutions, the non-cooperative approach is simple to apply, in which the OBUs share no information and response to any data request from other OBUs. However, since the transmissions are not coordinated, the non-cooperative approach might be inefficient due to the collisions and repeated transmissions. Therefore, we consider to introduce cooperation among OBUs, i.e., the OBUs can share some information to make the P2P approach more efficient. Particularly, we adopt the coalitional graph game model as first introduced by Myerson in \cite{Myerson-1977}. In a coalitional graph game, the players try to maximize their individual payoffs, which are related to the specific graph that interconnects them. This model has recently been used in many graph-based communication problems \cite{SHDHB-2009}. In \cite{SHDH-2008}, the author investigated the problem of the formation of an uplink tree structure in the IEEE 802.16j standard with the coalitional graph game model, in which the relay stations form a directed tree graph to improve their utility considering both packet success rate and link maintenance cost. In \cite{AJM-2007}, the author considered a network formation game where the nodes wish to send traffic to each other and a form of the myopic best response dynamics is proposed. In the proposed PCD problem, we introduce a coalitional graph game model considering content request, peer location, channel capacity, and network topology, and a coalition formation algorithm is proposed to implement the P2P approach based on this model.

Also, we utilize cognitive radio (CR) for V2V transmissions in the proposed approach. In VANETs, V2V links might be blocked by the factors such as deep channel fadings and severe data collisions. By exploiting CR, it is possible to utilize other available channels with better channel conditions, which in result, increases the sum transmission rate and reduces data collisions in the network. Therefore, CR is better for QoS for delay sensitive applications. Moveover, the vehicles can easily provide sufficient power and space that are especially needed by CR devices. And in vehicular environments, especially in suburban highways, the spectrum is relatively clean and there are plenty of spectrum holes that can by utilized by CR. Thus, CR is beneficial in VANETs. Considering all the benefits, we utilize CR for the vehicle-to-vehicle (V2V) transmissions in the P2P approach.

The main contribution of this paper is to introduce a coalitional graph game model to the popular content distribution problem in VANETs. Based on the game theory model, a distributed coalition formation algorithm is proposed, in which the OBUs self-organize into coalitions to coordinate their V2V and V2I transmissions. Simulation results show that, using the proposed algorithm, the maximal total throughput is increased by $133\%$ and $250\%$, respectively, relative to the non-cooperative approach and the pure broadcasting scheme. And the total possessed packets are increased by $25\%$ and $218\%$, respectively, relative to the non-cooperative approach and the pure broadcasting scheme.

The rest of this paper is organized as follows. Section \uppercase\expandafter{\romannumeral2} provides the system model and the non-cooperative approach. In Section \uppercase\expandafter{\romannumeral3}, we formulate the graph-based utility function and present a coalitional graph game for modeling the transmitting behaviors of OBUs. In Section \uppercase\expandafter{\romannumeral4}, we exposes the properties of the coalitional graph game as well as the dynamics algorithm for forming a best-response network graph. In Section \uppercase\expandafter{\romannumeral5} simulation results and analysis are presented, and in Section \uppercase\expandafter{\romannumeral6} we conclude the paper.

\section{System Model}%

\subsection{Network Model}

Consider a VANET consisting of $N$ OBUs driving through the coverage of a RSU. A popular content, which has been equally divided into $M$ packets with the size of each packet $s$, is broadcasted at an authorized frequency band (e.g. the $75$MHz bandwidth in the 5.9GHz band for IEEE 802.11p \cite{802}) from the RSU to the OBUs inside the coverage. We suppose the V2R transmission time is not sufficient for any OBU to download the entire file, but a few packets. Let $\mathcal{N}$, $\mathcal{M}$ and $\mathcal{M}_i$ denote the set of OBUs, and the set of packets, and the set of packets possessed by OBU $i$, respectively.

For downloading of the entire file, a P2P network is built among OBUs by unauthorized channels using cognitive radio. We suppose there are $K$ such channels with primary traffic modeled as $K$ independent Poisson processes with the same arriving rate $\lambda$ per time slot. In each slot, every OBU in the network transmits, or not, to another OBU through an empty channel. The OBUs keep exchanging their possessed packets until every OBU in the network achieves the entire file, formally, $\mathcal{M}_i = \mathcal{M}, \forall i \in \mathcal{N}$. Let $\mathcal{K}$ and $\mathcal{K}_i$ denote the set of unauthorized channels and the set of unauthorized channels sensed by OBU $i$, respectively.

\begin{figure}[!t]
\centering
\includegraphics[width=4.2in]{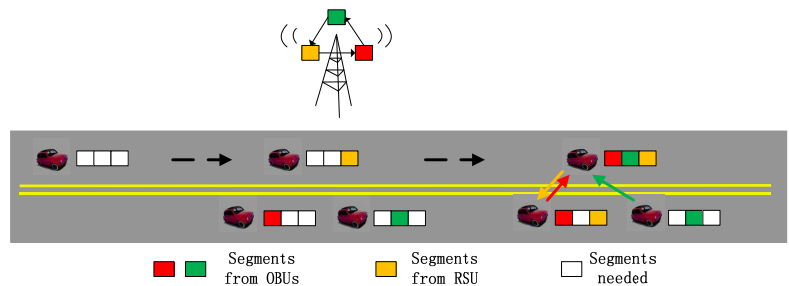}
\caption{System model of popular content distribution in cognitive radio ad hoc VANETs.} \label{system_model}
\end{figure}

As shown in Fig.~\ref{system_model}, the evolution of the popular file in an OBU is illustrated. When an OBU arrives in the range of the RSU, it immediately receives the broadcasted packets from the RSU. As getting out of range, it starts the P2P transmissions with other OBUs in unauthorized channels using cognitive radio. After finite P2P transmissions, the popular content can be entirely obtained by all OBUs in the network.

\subsection{Channel Model}

In this paper, we suppose all OBUs and the RSU are equipped with single omnidirectional antenna. The RSU periodically broadcasts the popular file at an authorized frequency with data rate $R_0$. For the V2R channels between the RSU and the OBUs inside its coverage, we adopt Rayleigh model for small-scale fading and a path loss model with the path loss exponent equal to $4$. For simplicity without loss of generality, we do not consider the shadowing by other vehicles. Also, we assume in the same slot the channel is unchanged. Thus, the gain of the V2R channel between the RSU and any OBU $i$, denoted by $h_i$, is given by
\begin{equation} \label{V2RGain}
h_i = \alpha {d_i}^{-2},
\end{equation}
where $\alpha$ is a complex Gaussian random variable with unit variance and zero mean, and $d_i$ is the distance between the RSU and OBU $i$. Therefore, the capacity of the V2R channel between the RSU and OBU $i$, denoted by $c_i$, is given by
\begin{equation} \label{V2RCapacity}
c_i = W \log_2{\left(1+ \beta \left| {h_i} \right|^2 \right)},
\end{equation}
where $W$ is the bandwidth authorized for V2R communications and $\beta$ a scale factor representing the transmit power of the RSU. In each slot, we assume OBU $i$ can receive useful data from the RSU if and only if $c_i > R_0$.

For the V2V channels, we suppose the transmitting signal from any OBU $i$ can only be received by its ``neighbors" (the OBUs with a line of sight (LOS) to OBU $i$) \cite{MMKTPBZKC-2011}, denoted by $\mathcal{N}_i$. For those channels with a LOS, we adopt a similar model as the V2R channels. Thus, the channel gain $h^k_{i,j}$ between OBU $i$ and OBU $j \ne i$ in the $k$-th unauthorized channel is given by
\begin{equation} \label{V2VGain}
h^k_{i,j}=
\left\{ \begin{gathered}
\alpha {d_{i,j}}^{-2}, \mbox{a LOS exists}, \hfill \\
0~~~~~~~,\mbox{otherwise}, \hfill \\
\end{gathered} \right.
\end{equation}
where $d_{i,j}$ is the distance between OBU $i$ and OBU $j$, and $k = 1,2,\ldots,K$. Therefore, the corresponding capacity, denoted by $c^k_{i,j}$, is given by
\begin{equation} \label{V2VCapacity}
c^k_{i,j} = W_k \log_2{\left(1+ \beta_i \left| {h^k_{i,j}} \right|^2 \right)}, k = 1,2,\ldots,K,
\end{equation}
where $W_k$ is the bandwidth of the $k$-th channel, $\beta_i$ a scale factor representing the transmit power at OBU $i$. For simplicity without loss of of generality, we assume $W_1 = W_2 = \ldots, W_K = W'$ and $\beta_1 = \beta_2 = \ldots, \beta_K = \beta'$.

\subsection{Non-cooperative Approach}

In the non-cooperative manner, the P2P transmissions do not coordinate with each other. For any OBU $i \in \mathcal{N}$, its packet request is broadcasted to its ``neighbors" at the beginning of each slot, and the first responding OBU, e.g., OBU $j$, transmits the requested packets to OBU $i$. For finding available channels, OBU $j$ randomly senses $K_j < K$ unauthorized channels, the set of which is denoted by $\mathcal{K}_j \subseteq \mathcal{K}$, and for each channel with a constant time $\tau$. If the period of a slot is $T$, the transmitting time left is then $T-K_j\tau$, and the corresponding rate is given by
\begin{equation} \label{V2VRate}
R^k_{j,i} = \left(\frac{T - K_j\tau}{T}\right)c^k_{j,i}, k \in \mathcal{K}_j.
\end{equation}
The channel with the largest transmission rate is selected by OBU $j$, which is given by
\begin{equation} \label{V2VBestRate}
R^*_{j,i} = \max \limits_{k \in \mathcal{K}_j} R^k_{j,i} = \left(\frac{T - K_j\tau}{T}\right) \times \max \limits_{k \in \mathcal{K}_j} c^k_{j,i}.
\end{equation}

We suppose the relation between $K_j$ and the maximum channel capacity $c^*_{j,i}$ can be given by a function $f(K_j)$ in the average sense. The transmission rate in (\ref{V2VBestRate}) is then rewritten as
\begin{equation} \label{V2VPraticalRate}
R^*_{j,i} = \left(\frac{T - K_j\tau}{T}\right) f(K_j).
\end{equation}
To maximize $R^*_{j,i}$, we have $K_j$ must satisfy
\begin{equation} \label{KjDecide1}
\frac{d R^*_{j,i}}{d K_j} = 0. \Rightarrow K_j + \frac{f(K_j)}{f'(K_j)} = \frac{T}{\tau}.
\end{equation}
We further suppose $f(K_j) = A \ln{(K_j+1)}$ and $T = K\tau$, then, (\ref{KjDecide1}) is rewritten as
\begin{equation} \label{KjDecide2}
(K_j+1) \ln{(K_j+1)} + K_j = K.
\end{equation}
Equation (\ref{KjDecide2}) can be easily solved by numerical solutions and the result can be seen as a reference value for $K_j$. We adopt a logarithmic form as the expression of $f(K_j)$ for the following reasons: (1) $f(K_j)$ must be an increasing function, since an extra channel can only increase, or at least, maintains the maximum capacity in average sense, (2) $f(K_j)$ must be a concave function, since the contribution of an extra channel will be mitigated by the increasing channel number.

In the above analysis, we do not consider any data collisions, which, however, are inevitable. As we noted, the transmitting signal from any OBU $i$ can only be received by its ``neighbors". Thus, the signal as well as the interference are both confined in the ``neighbors". Considering the potential collisions with primary users (the users with spectrum license) and other OBUs, the successful transmitting conditions are: (1) a LOS exists between OBU $i$ and OBU $j$, (2) no primary traffic exists in the $k$-th channel, (3) no ``neighbors" of OBU $i$ transmit in the $k$-th channel. In fact, the influence of collisions has been involved in factor $A$ when we suppose $f(K_j) = A \ln{K_j}$, which makes no difference in the result.

\section{Coalitional Graph Game Model}%

For the proposed PCD problem, we have provided a non-cooperative solution, in which the sensing-throughput tradeoff has been locally optimized. However, due to the random establishment of V2V links, the OBUs may be connected to inefficient ``neighbors" or even with no ``neighbor" to be connected to. Thus, the entire network may suffer from low data throughput, or equivalently, high transmission delay. In this section, we introduce a coalitional graph game model to coordinate the V2V links between different OBUs and also the V2R links from the RSU to the OBUs. In this model, each OBU decides to connect to or be connected to other OBUs in order to maximize its own utility that takes into account data throughput as well as link maintaining cost. The result of the interactions among the OBUs is a directed graph $G(\mathcal{V},\mathcal{E})$ with $\mathcal{V}$ denoting the set of all nodes (the OBUs and the RSU) and $\mathcal{E}$ denoting the set of all edges (V2V links and V2R links). For any $i,j \in \mathcal{V}$, we say the link from $i$ to $j$ exists, if $e_{i,j} \in \mathcal{E}$.

\subsection{Utility Function}

For any OBU $i \in \mathcal{N}$, we suppose positive utilities can be extracted from both the effective packets received from and transmitted to other OBUs, which depends on the current links associated with OBU $i$. Thus, the utility function is a graph-based function, which is denoted by $U_i(G)$. Due to the single antenna, the maximum number of connections from or to OBU $i$ is confined to $1$, which implies the out-degree $\lambda^{out}_i$ and in-degree $\lambda^{in}_i$ of node $i$ in graph $G$ satisfying $0 \le \lambda^{out}_i \le 1, 0 \le \lambda^{in}_i \le 1$. Using the broadcasting channel, the RSU can connect to as many OBUs as possible while no OBU can connect to it, which implies out-degree $\lambda^{out}_{RSU}$ and in-degree $\lambda^{in}_{RSU}$ of the RSU in graph $G$ satisfying $\lambda^{out}_{RSU} \ge 0, \lambda^{in}_{RSU} = 0$.

The corresponding utility of the received packets, denoted by $U^{in}_i(G)$, is assumed to be proportional to the number of received packets, which is given by
\begin{equation} \label{UtilityIn}
U^{in}_i(G)=
\left\{ \begin{gathered}
\gamma_{in} P_{j,i} \min\left( \frac{R^*_{j,i} T}{s}, \left|\left( \mathcal{M} \backslash \mathcal{M}_i \right) \cap \mathcal{M}_j \right| \right), \lambda^{in}_i = 1 ~\&~ e_{j,i} \in \mathcal{E}, \hfill \\
\gamma_{in}\frac{c_i T}{s}~~~~~~~~~~~~~~~~~~~~~~~~~~~~~~~~~~~~~~~, \lambda^{in}_i = 1 ~\&~ e_{RSU,i} \in \mathcal{E}, \hfill \\
0~~~~~~~~~~~~~~~~~~~~~~~~~~~~~~~~~~~~~~~~~~~~~~,\lambda^{in}_i = 0, \hfill \\
\end{gathered} \right.
\end{equation}
where $\gamma_{in} >0$ is a pricing factor and $P_{j,i}$ is the probability of successful transmission from OBU $j$ to OBU $i$, the expression of which will be given in the following part of the section.

The corresponding utility of the transmitted packets, denoted by $U^{out}_i(G)$, is assumed to be proportional to the number of transmitted packets, which is given by
\begin{equation} \label{UtilityOut}
U^{out}_i(G)=
\left\{ \begin{gathered}
\gamma_{out} P_{i,j} \min\left( \frac{R^*_{i,j} T}{s}, \left|\left( \mathcal{M} \backslash \mathcal{M}_j \right) \cap \mathcal{M}_i \right| \right), \lambda^{out}_i = 1 ~\&~ e_{i,j} \in \mathcal{E}, \hfill \\
0~~~~~~~~~~~~~~~~~~~~~~~~~~~~~~~~~~~~~~~~~~~~~~,\lambda^{out}_i = 0, \hfill \\
\end{gathered} \right.
\end{equation}
where $\gamma_{out}>\gamma_{in}$ is a pricing factor and $P_{i,j}$ is the probability of successful transmission from OBU $i$ to OBU $j$, the expression of which will be given in the following part of the section. The main driver behind the transmitting benefit is that the importance of the role of OBU $i$ increases if OBU $i$ serves more OBUs, and, thus, its utility should increase.

When OBU $i$ transmits or receives data, certain channels are occupied, which increases the probability of data collisions. The cost of potential collisions is proportional to the number of OBUs that has been interfered. For the link from OBU $i$, the interference is confined in OBU $i$'s ``neighbors" $\mathcal{N}_i$. For the link from OBU $j$ to OBU $i$, the interference is confined in OBU $j$'s ``neighbors" $\mathcal{N}_j$. Thus, the cost function, denoted by $C_i(\lambda^{out}_i,\lambda^{in}_i)$, is given by
\begin{equation} \label{Cost}
C_i(\lambda^{out}_i,\lambda^{in}_i)=
\left\{ \begin{gathered}
\gamma_{cost} \lambda^{out}_i \left|\mathcal{N}_i\right| + \gamma_{cost} \lambda^{in}_i \left|\mathcal{N}_j\right| ,\lambda^{in}_i = 1 ~\&~ e_{j,i} \in \mathcal{E}, \hfill \\
\gamma_{cost} \lambda^{out}_i \left|\mathcal{N}_i\right|~~~~~~~~~~~~~~~~~~, \mbox{otherwise}, \hfill \\
\end{gathered} \right.
\end{equation}
where $\gamma_{cost}>0$ is a pricing factor.

In summary, given a transmission graph $G(\mathcal{V},\mathcal{E})$, the cost and benefit of OBU $i \in \mathcal{N}$ is captured by the following utility function
\begin{equation} \label{Utility}
U_i(G) = U^{in}_i(G) + U^{out}_i(G) - C_i(\lambda^{out}_i,\lambda^{in}_i).
\end{equation}

\subsection{Collisions}

For any given unauthorized channel $k \in \mathcal{K}$ and any OBU $i \in \mathcal{N}$, the probability of miss (i.e., probability of missing the detection of the primary traffic) and false alarm (i.e., probability of the false detection of the primary traffic) are denoted by $P^i_m(k)$ and $P^i_f(k)$, respectively. For simplicity without loss of generality, we suppose the sensing devices in all OBUs have the same performance for any channels, which implies $P^i_m(k) = P_m, P^i_f(k) = P_f, \forall i \in \mathcal{N}, k \in \mathcal{K}$. As we noted, the primary traffic in any channel $k$ is modeled as a Poisson process with parameter $\lambda$. Thus, the probability that no primary traffic occupies channel $k$ is denoted by $P_0 = e^{-\lambda}$.

We consider the probability of successful transmission from OBU $i$ to OBU $j$, denoted by $P_{i,j}$. For a successful transmission, the link from OBU $i$ to OBU $j$ should not be interfered by other OBUs, which implies that none of OBU $j$'s ``neighbors" is transmitting at the same channel. In the average sense, there are $P_0K$ channels available for V2V transmission and OBU $i$ occupies one of them. Thus, the probability that OBU $j$'s neighbors do not transmit at the same channel is given by
\begin{equation} \label{ProbabiliySU}
P^a_{i,j} = \left(\frac{KP_0-1}{KP_0}\right)^{\left|\mathcal{N}_i\right|}.
\end{equation}
We denote by $\mathcal{H}_1$ the hypothesis that the unauthorized channel is occupied by a primary user in real, and $\mathcal{H}_0$ the alternative hypothesis. Also, we denote by $\mathcal{H}_1'$ the hypothesis that the sensing result shows the unauthorized channel is occupied, and $\mathcal{H}_0'$ the alternative hypothesis. Thus, the possibility that the empty decision of the unauthorized channel is correct can be expressed by:
\begin{align} \label{ProbabiliyPU1}
P\left({H_0}|{H_0'}\right) =  \frac{P(H_0)P(H_0'|H_0)}{P(H_0'|H_0)P(H_0) + P(H_0'|H_1)P(H_1)},
\end{align}
where $P(H_0) = P_0$, $P(H_1) = 1 - P_0$, $P(H_0'|H_1) = P_m$, and $P(H_0'|H_0) = 1 - P_f$. Note that (\ref{ProbabiliyPU1}) is also the probability that the transmission from OBU $i$ to OBU $j$ does not collide with primary traffic, we rewrite this probability as
\begin{equation} \label{ProbabiliyPU2}
P^b_{i,j} = \frac{P_0(1-P_f)}{P_0(1-P_f) + (1-P_0)P_m}.
\end{equation}
Thus, the probability of successful transmission from OBU $i$ to OBU $j$ is given by
\begin{equation} \label{ProbabiliySuccess}
P_{i,j} = P^a_{i,j} P^b_{i,j} = \left(\frac{KP_0-1}{KP_0}\right)^{\left|\mathcal{N}_i\right|} \left( \frac{P_0(1-P_f)}{P_0(1-P_f) + (1-P_0)P_m} \right),
\end{equation}
which completes the expressions of $U^{in}_i(G)$, $U^{out}_i(G)$ in (\ref{UtilityIn}), (\ref{UtilityOut}), respectively.

\section{Network Formation Algorithm}%

In this section, we propose a myopic dynamics algorithm for the coalitional graph game, which results in a directed graph that coordinates the transmissions in the network. \emph{Myopic dynamics} refer to the property of the dynamics that at any given round, nodes update their strategic decisions only to optimize their current utility, in contrast to dynamics that consider some long-term objective \cite{AJM-2007}. We show that the proposed algorithm is distributed and localized algorithm that results in a \emph{local Nash} network, which adapts to the environmental changes.

\subsection{Network Formation Algorithm}

The proposed network formation algorithm is distributively carried out by each OBU in the network. Given the transmission graph $G(\mathcal{V},\mathcal{E})$, the available strategies any OBU $i \in \mathcal{N}$ are classified as follows:
\begin{enumerate}
    \item Offer OBU $j \ne i$ a new link $e_{i,j}$, if $\lambda^{out}_i = 0$.
    \item Break the link $e_{i,k}$, if $\lambda^{out}_i = 1, e_{i,k} \in \mathcal{E}$.
    \item Accept $j$'s (OBU $j$ or the RSU) request for the link $e_{j,i}$, if $\lambda^{in}_i = 0$.
    \item Break the link $e_{k,i}$, if $\lambda^{in}_i = 1, e_{k,i} \in \mathcal{E}$.
    \item Combinations of item $1\sim4$.
\end{enumerate}

Formally, denote $(a_i,b_i)$ as the state of OBU $i$, where $a_i \in \mathcal{N}\cup\{RSU\}$ is the node that transmits to OBU $i$ ($a_i = i$ refers to $\lambda^{in}_i = 0$), and $b_i \in \mathcal{N}$ is the OBU that OBU $i$ transmits to ($b_i = i$ refers to $\lambda^{out}_i = 0$). Thus, the state space of OBU $i$ is given by $\{(a_i,b_i) ~|~ a_i \in \mathcal{N}\cup\{RSU\}, b_i \in \mathcal{N}\}$. By carefully inspecting the strategies listed above, we find that those strategies correspond to their consequent states of OBU $i$. Thus, the strategy space $S_i$ is also the state space of OBU $i$. Formally, denote $S_i = \{(a_i,b_i) ~|~ a_i \in \mathcal{N}\cup\{RSU\}, b_i \in \mathcal{N}\}$ as the strategy space of OBU $i$ and $s_i \in S_i$ as the strategy of OBU $i$.

Note that the proposed algorithm focuses on maximizing the current utility $U_i$ of OBU $i$, some of the strategies in $S_i$ may be infeasible for reducing the utilities of corresponding OBUs. For strategies that only involves in breaking down old links (e.g., item 2 and item 4), the utility $U_i$ of OBU $i$ should be improved, or at least maintained the same. For strategies that involves in forming new links (e.g., item 1 and item 3), both the utility $U_i$ of OBU $i$ and the utility $U_j$ of OBU $j$ should be improved, or at least maintained the same. Thus, we give the following definition.

\begin{definition}\label{def1}
A local strategy $s_i = ({a_i}',{b_i}') \in S_i$ is a \emph{feasible local strategy} for OBU $i \in \mathcal{N}$ with state $(a_i,b_i)$, if and only if: (1) $U_i(G') \ge U_i(G)$, (2) $U_{{a_i}'}(G') \ge U_{{a_i}'}(G)$ for ${a_i}' \ne i, {a_i}' \ne a_i, {a_i}' \in \mathcal{N}$, (3) $U_{{b_i}'}(G') \ge U_{{b_i}'}(G)$ for ${b_i}' \ne i, {b_i}' \ne b_i$, where $G(\mathcal{V},\mathcal{E})$ is the current transmission graph and $G'(\mathcal{V},\mathcal{E}')$ is the consequent transmission graph by strategy $s_i$. Denote $F_i \subseteq S_i$ as the set of feasible strategies for OBU $i$.
\end{definition}

Denote $G_{s_i,{\bf{s}}_{-i}}$ as the graph $G$ formed when OBU $i$ plays a feasible strategy $s_i \in F_i$ while all other OBUs maintain their vector of strategies ${\bf{s}}_{-i} = [s_1,\ldots,s_{i-1},s_{i+1},\ldots,s_N]$. We define the local best response as follows \cite{DKTT-2008}.

\begin{definition}\label{def2}
A \emph{feasible local strategy} $s_i \in F_i$ is a \emph{local best response} for OBU $i \in \mathcal{N}$ if $U_i(G_{s^*_i,{\bf{s}}_{-i}}) \ge  U_i(G_{s_i,{\bf{s}}_{-i}}), \forall s_i \in F_i$. Thus, the \emph{local best response} for OBU $i$ is to select the links that maximizes its utility given that the other OBUs maintain their vector of strategies.
\end{definition}

We assume that each OBU is myopic, in the sense that the OBUs aim at improving their utilities considering only the current state of the network. Several models for \emph{myopic dynamics} have been considered in the literature \cite{DKTT-2008,JW-2002,DW-2005}. Here, we propose a myopic dynamics algorithm composed of indefinite number of rounds, where in each round all OBUs in $\mathcal{N}$ update their \emph{local best responses} by random priority. If the algorithm converges to a final graph $G^*(\mathcal{V},{\mathcal{E}}^*)$ after finite rounds, we adopt $G^*$ to coordinate the transmissions in the network, that is, node $i$ (OBU $i$ or the RSU) transmits to OBU $j$ if and only if $e_{i,j} \in {\mathcal{E}}^*$.

\subsection{Convergency and Stability}

Having an analytical proof for the convergency of network formation games, especially with practical utilities and discrete network formation strategies, is difficult \cite{DW-2005,BO-1999}. In fact, in wireless applications \cite{SMG-2001,NH-2007,BAS-2003}, it is common to propose best-response algorithms without any analytical proof of convergency, since such algorithms can, in most cases, converge.

If the proposed algorithm converges to a final graph $G^*$, we define the following concept \cite{DKTT-2008}:

\begin{definition}
A network graph $G$, in which no node $i$ can improve its utility by a unilateral change in its \emph{feasible local strategy} $s_i \in F_i$, is a \emph{local Nash} network.
\end{definition}

\begin{lemma}
The final graph $G^*$ resulting from the proposed algorithm is a local Nash network.
\end{lemma}

\begin{IEEEproof}
When the proposed algorithm converges to a final graph $G^*(\mathcal{V},\mathcal{E}^*)$, any strategy $s_i$ that OBU $i \in \mathcal{N}$ maintains must be the \emph{local best response} $s^*_i$. Thus, we have $U_i(G_{s^*_i,{\bf{s}}_{-i}}) \ge  U_i(G_{s_i,{\bf{s}}_{-i}}), \forall s_i \in F_i$ from the definition of \emph{local best response}. In consequence, there is no OBU $i \in \mathcal{N}$ can improve its utility by a unilateral change in its \emph{feasible local strategy} $s_i \in F_i$. Hence, the final graph $G^*$ is a \emph{local Nash} network.
\end{IEEEproof}

In the case of non-convergence, the proposed algorithm may cycle between a number of networks. In order to avoid such undesirable cycles, one can introduce additional constraints on the strategies such as allowing the nodes to select their strategies, not only based on the current network, but also on the history of moves or strategies taken by the other nodes. In \cite{SHBDH-2011}, the author allows the nodes to observe the visited networks during the occurrence of network formation. By setting up an upper bound for the occurrence number of a particular network, those potential cycles could be broken and the proposed game can converge. This constraint need every node be aware of the topology of the entire network, which is not a big problem in \cite{SHBDH-2011} since the nodes are relay stations with fixed locations and high transmitting power. However, in our scenario, where the nodes are OBUs with high mobility and limited transmitting power, it is difficult for every OBU to achieve the entire topology within a slot-level period. Therefore, in our proposed algorithm, we allow the OBUs to count for each used strategy and set an upper bound for it. Formally, we define a history function $h^t_i(s_i)$ which represents, for every used strategy $s_i \in S_i$ (note the strategy space is now $S_i$), the number of times this strategy was used by OBU $i$ in the past $t$ iterations. Further, we define a threshold $\sigma$ (positive integer) for $h^t_i(s_i)$ above which OBU $i$ is no longer interested in adopting this strategy, since it may lead to a cyclic behavior. Thus, at any iteration $t+1$, there is an additional step for any OBU $i\in \mathcal{N}$, which is to update the set of its feasible strategies by
\begin{align} \label{UpdateFeasible}
{F_i}' = F_i \backslash \{s_i ~|~ h^t_i(s_i) \ge \sigma\},
\end{align}
where $F_i$ is the original set of feasible strategies defined in definition \ref{def1}. The proposed algorithm with strategy constraint is summarized in Table \ref{NetworkFormation}.

By adding the strategy constraint to the algorithm, the coalitional graph game will converge after finite iterations, which is guaranteed by the following theorem:

\begin{theorem}
Given any initial network graph $G_0$, there exist a positive integer $T$, that the proposed algorithm with strategy constraint will converge after $T$ iterations.
\end{theorem}

\begin{IEEEproof}
Suppose the algorithm does not converge after $T$ iterations. In this regard, denoting by $G_t$ the graph reached at the end of any iteration $t$, the proposed algorithm consists of a graph sequence such as the following
\begin{align} \label{GraphSequence}
G_0 \rightarrow G_1 \rightarrow G_2 \rightarrow \ldots \rightarrow G_t \rightarrow  \ldots  \rightarrow G_T.
\end{align}
According to the pigeonhole principle, there exists a graph $G^x(\mathcal{V},\mathcal{E}^x)$ that occurs more than $T / \left|\mathcal{G}\right|$ times, where $\mathcal{G} = \{G(\mathcal{V},\mathcal{E}) ~|~ \forall i\in \mathcal{N}, \lambda^{in}_i \le 1, \lambda^{out}_i \le 1\}$ denotes the set of all possible network graphs. Again according to the pigeonhole principle, there exists an OBU $x \in \mathcal{N}$ that uses the strategy $s_x = (a_x,b_x)$ more than $T / (\left|\mathcal{G}\right|N)$ times, where $e_{a_x,x}, e_{x,b_x} \in \mathcal{E}^x$. We suppose $T = \left|\mathcal{G}\right|N(\sigma + 1)$. Then, after $T$ iterations, there exits an OBU $x \in \mathcal{N}$ and one of its strategy $s_x \in S_x$, where the corresponding history function satisfying
\begin{align} \label{contradiction}
h^T(s_x) > \frac{T}{N\left|\mathcal{G}\right|} \Rightarrow h^T(s_x) > \sigma,
\end{align}
which contradicts to our constraint. Thus, the proposed algorithm with strategy constraint must converge after $T$ iterations, where $T$ is a finite number.
\end{IEEEproof}

\subsection{Scalability and Adaptation to Environmental Changes}

The proposed algorithm in Table \ref{NetworkFormation} adopts a distributed approach to form a network graph, which coordinates all transmissions for the proposed PCD application. By carefully inspecting the utility function $U_i(G)$ in (\ref{Utility}), we find that, for calculating $U_i(G)$, the topology of transmission graph $G$ is not essential. All information OBU $i$ needs can be achieved from OBU $i$'s ``neighbors" $\mathcal{N}_i$, which includes channel conditions, the possessed packets of ``neighbors", the number of a ``neighbor's neighbors". Moreover, the strategy constraint we introduce can also be performed in a localized manner. Hence, the proposed algorithm is a localized approach with overhead only between ``neighbors", which is, therefore, a scalable algorithm.

In the proposed PCD problem, the myopic dynamics algorithm is repeated periodically every slot, which allows the OBUs to take autonomous decisions to update the transmission topology adapting to any environmental changes. As the slot is chosen to be shorter, the proposed algorithm is played more often, allowing a better adaptability. Moreover, as the environmental changes mitigate, we can expect the change of the final transmission graph also mitigates, which implies less complexity in calculation since the transmission graph will be inherited by the next slot.

In the proposed algorithm, the OBU selects its transmitting channel individually without any coordination, which may, when the network is dense, introduce considerable interference. To avoid the potential collisions, interference management can be introduced, e.g., the cooperative approach in \cite{SHZHBP-2012}, where each node assigns different weights on the channels and cooperatively sort their channels, in a manner to reduce interference as much as possible. We do not introduce such management for the following reasons:
\begin{enumerate}
    \item The proposed PCD problem is in highway scenarios, where the spectrum is relatively clean and the traffic flow is not heavy. Thus, the interference management may be unnecessary.
    \item In order to implement the cooperative interference management, each node needs to share its information with the entire network, which ruins the localized property of our algorithm.
\end{enumerate}

\section{Simulation Results}%

In this section, the performance of the proposed algorithm in Table \ref{NetworkFormation} is simulated in various environmental conditions, compared with the pure broadcasting scheme, the non-cooperative approach, and the optimal solution. Here, the pure broadcasting is a scheme in which no V2V transmission is allowed and all the OBUs can only receive from the RSU. The pure broadcasting scheme can be seen as a lower bound for evaluating all possible P2P approaches. Also, we give an upper bound with the optimal solution, in which the overall information has been considered and an optimal solution has been derived by enumeration. The mobility model and system parameters are presented in the following subsection.

\subsection{Mobility Model and System Parameters}

The mobility model we use is similar to the Freeway Mobility Model (FMM) proposed in \cite{MPGW-2006}, which is well accepted for modeling the traffic in highway scenarios. In FFM, the simulation area includes many multiple lane freeways without intersections. At the beginning of the simulation, the vehicles are randomly placed in the lanes, and move at history-based speeds. The vehicles randomly accelerate or decelerate with the security distance $d_{min}>0$ maintained between two subsequent vehicles in the same lane and no change of lanes is allowed.

In our scenario, the map has been simplified to a one-way traffic road with double lanes as shown in Fig.~\ref{system_model}. All the OBUs independently choose to speed up or slow down by probability $p$ and acceleration $a>0$. The velocity of any OBU $i\in \mathcal{N}$ is limited by $v_{min} \le v_i(t) \le v_{max}$ for all time. To better reflect the changing topology of VANETs, we decide to allow the change of lanes when a vehicle is overtaking, as long as the security distance is maintained. Also, to prevent the vehicles from being widely scattered, we also give an upper bound $d_{max}$ for the distance between any two subsequent vehicles in the same lane. The overall constraints in our mobility model are listed as follows:
\begin{enumerate}
    \item The OBUs are randomly placed on both lanes in an area with length $L$ that is $D$ away form the RSU when the simulation begins.
    \item The initial speed of OBU $i \in \mathcal{N}$, denoted by $v_i(0)$, is randomly given in $[v_{min}, v_{max}]$.
    \item The speed of OBU $i \in \mathcal{N}$ satisfies:
            \begin{align} \label{Velocity}
            {v_i}(t+1) = \left\{
            \begin{array}{rl}
            {v_i}(t)~~~~~~~~~~~~~~~~~~~~,&1-2p, \\
            \min {\left({v_i}(t)+a,v_{max}\right)},&p, \\
            \max {\left({v_i}(t)-a,v_{min}\right)},&p, \\
            \end{array}
            \right.
            \end{align}
        where $p$ is the probability of acceleration or deceleration.
    \item For any OBU $i\in \mathcal{N}$ with OBU $j_1$ ahead in the same lane and OBU $j_2$ ahead in the other lane, OBU $i$ switches to the other lane, if ${d_{i,j_1}}(t) \le d_{min} $ and ${d_{i,j_2}}(t) > d_{min}$, or OBU $i$ decelerates to $v_i(t+1) = v_{min}$, if ${d_{i,j_k}}(t) \le d_{min}, k=1,2$.
    \item For any OBU $i\in \mathcal{N}$ with OBU $j_1$ ahead in the same lane, OBU $i$ accelerates to $v_i(t+1) = v_{max}$, if ${d_{i,j_1}}(t) \ge d_{max}$.
\end{enumerate}

The parameters are taken from a general highway scenario as shown in Table \ref{Parameters}.

\subsection{Simulation Results}

\begin{figure}[!t]
\centering
\includegraphics[width=4.2in]{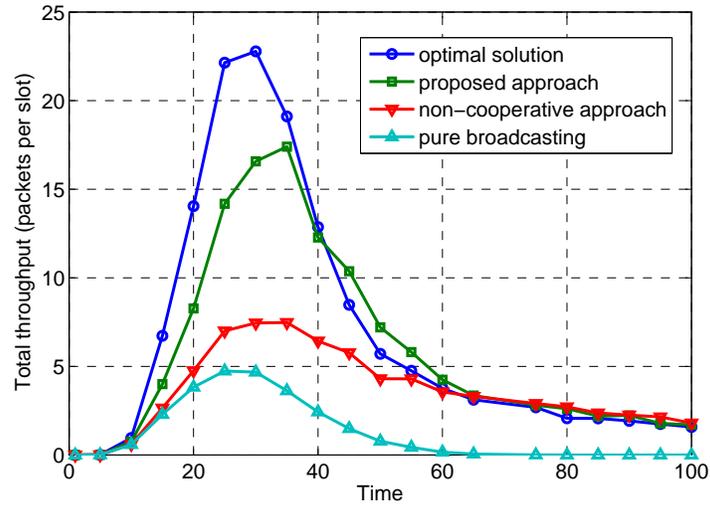}
\caption{Total throughput as a function of time for networks with $N=10, K=10$ and $ K'=4$.} \label{service_rate}
\end{figure}

In Fig.~\ref{service_rate}, we show the total throughput as a function of time for networks with $N=10,K=10$, and $K'=4$. Let $P(t) = \sum\nolimits_{i\in \mathcal{N}}{\left|\mathcal{M}_i\right|}$ denote the total possessed packets in the network. Then, the total throughput is given by $dP(t)/dt$, which represents the packets successfully delivered at current slot $t$. For pure broadcasting scheme, the total throughput increases as the OBUs enter the communication range of the RSU, and then decreases to zero as the OBUs leaves this area. By introducing V2V transmissions as in the proposed approach as well as the non-cooperative approach, the total throughput is largely increased. Even after the OBUs completely leaves the RSU area, the packets can still be exchanged among the OBUs and the total throughput is still above zero. Further, by introducing cooperation among OBUs, the proposed approach reduces the probability of collision and avoids repeated transmissions, relative to the non-cooperative approach, and thus, has a better performance in total throughput. In particular, Fig.~\ref{service_rate} shows that, using the proposed cooperative approach, the maximal total throughput is increased by $133\%$ and $250$, respectively, relative to the non-cooperative approach and pure broadcasting approach. Note that as more packets are delivered in the network, the OBUs become less willing to exchange data and the potential throughput for V2V transmissions decreases. The total throughput of optimal solution may fall below other methods, and our proposed approach may fall below the non-cooperative approach, especially seen from slot $50$ to slot $100$ in Fig.~\ref{service_rate}.

\begin{figure}[!t]
\centering
\includegraphics[width=4.2in]{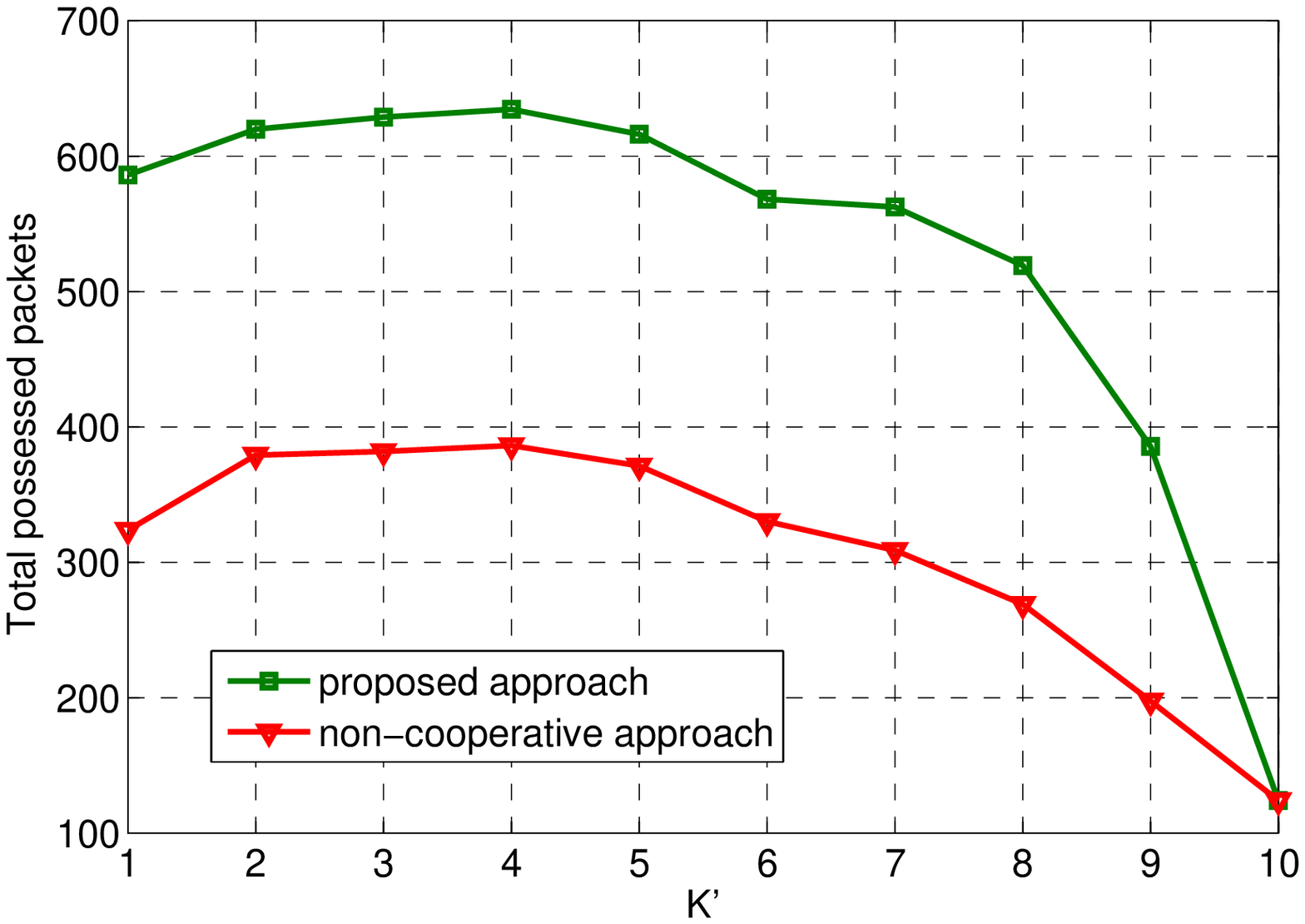}
\caption{Total possessed packets by the proposed approach and the non-cooperative approach at slot $t=100$, as a function of $K'$ for networks with $N=10$ and $K=10$.} \label{tradeoff}
\end{figure}

In Fig.~\ref{tradeoff}, we show the sensing-throughput tradeoff by setting different $K'$ from $1$ to $K$ in both the non-cooperative approach and the proposed algorithm. As we can see, the value of $K'$ with the best performance occurs at $K' = 4 \sim 5$, which coordinates to our reference value given in (\ref{KjDecide2}). Thus, the best spectrum sensing time $K'\tau$ can be individually decided by simple numerical solutions. Note that even $K'=K=10$, which means no time is left for V2V transmissions, the OBUs can still receive data from the RSU. The total possessed packets $P(t=100)$ is above zero for any $K'\le K$, as seen in Fig.~\ref{tradeoff}.

\begin{figure}[!t]
\centering
\includegraphics[width=4.2in]{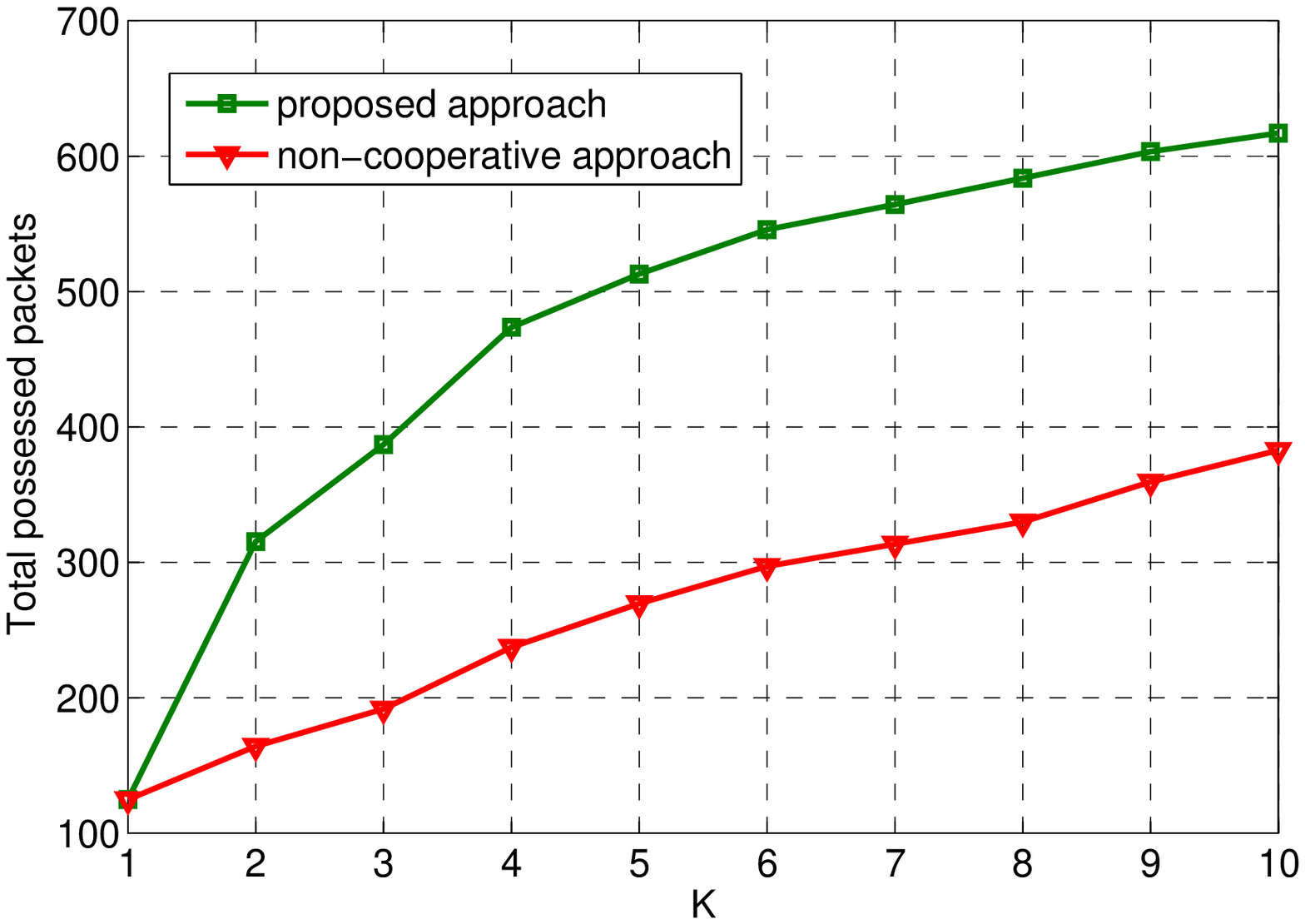}
\caption{Total possessed packets by the proposed approach and the non-cooperative approach at slot $t=100$, as a function of $K$ for networks with $N=10$ and $K'$ satisfying equation \ref{KjDecide2}.} \label{CR}
\end{figure}

In Fig.~\ref{CR}, we show the performance of the proposed algorithm with different numbers of cognitive radio channels for V2V transmissions. Here, the parameter $K'$, representing the number of sensed channels, has been chosen to satisfy equation (\ref{KjDecide2}). As we can see, the number of total possessed packets is highly effected by the cognitive spectrum. Therefore, we can expect a considerable performance improvement, when we apply CR for V2V transmissions, especially in highway scenarios with plenty of spectrum holes.

\begin{figure}[!t]
\centering
\includegraphics[width=4.2in]{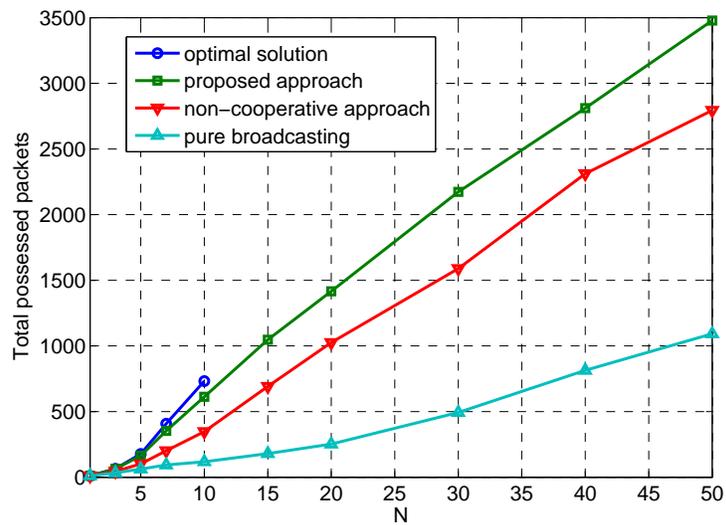}
\caption{Total possessed packets at slot $t=100$, as a function of $N$ for networks with $K=10$ and $K'=4$.} \label{scale}
\end{figure}

In Fig.~\ref{scale}, we show the performance of the proposed algorithm with different network sizes. Note that the total possessed packets increases linearly with the network size. We have the average possessed packets of each OBU is a stable constant with different network sizes, which implies the proposed algorithm is scalable and its performance is stable. This coordinates with our analysis in Section \uppercase\expandafter{\romannumeral4}.C, in which we point out that the overhead of the proposed approach is localized. As seen in Fig.~\ref{scale}, when the network is large ($N=50$), the total possessed packets of the proposed cooperative approach are increased by $25\%$ and $218\%$, respectively, relative to the non-cooperative approach and the pure broadcasting approach. When the network becomes smaller, the advantage declines gradually. In extremely sparse networks, as seen in Fig.~\ref{scale} when $N$ converges to $1$, there is few V2V transmission, so that all three P2P approaches degenerate to the pure broadcasting scheme. Note that the centralized optimal solution is shown for up to $N = 10$ since it is mathematically intractable for larger networks.

\begin{figure}[!t]
\centering
\includegraphics[width=4.2in]{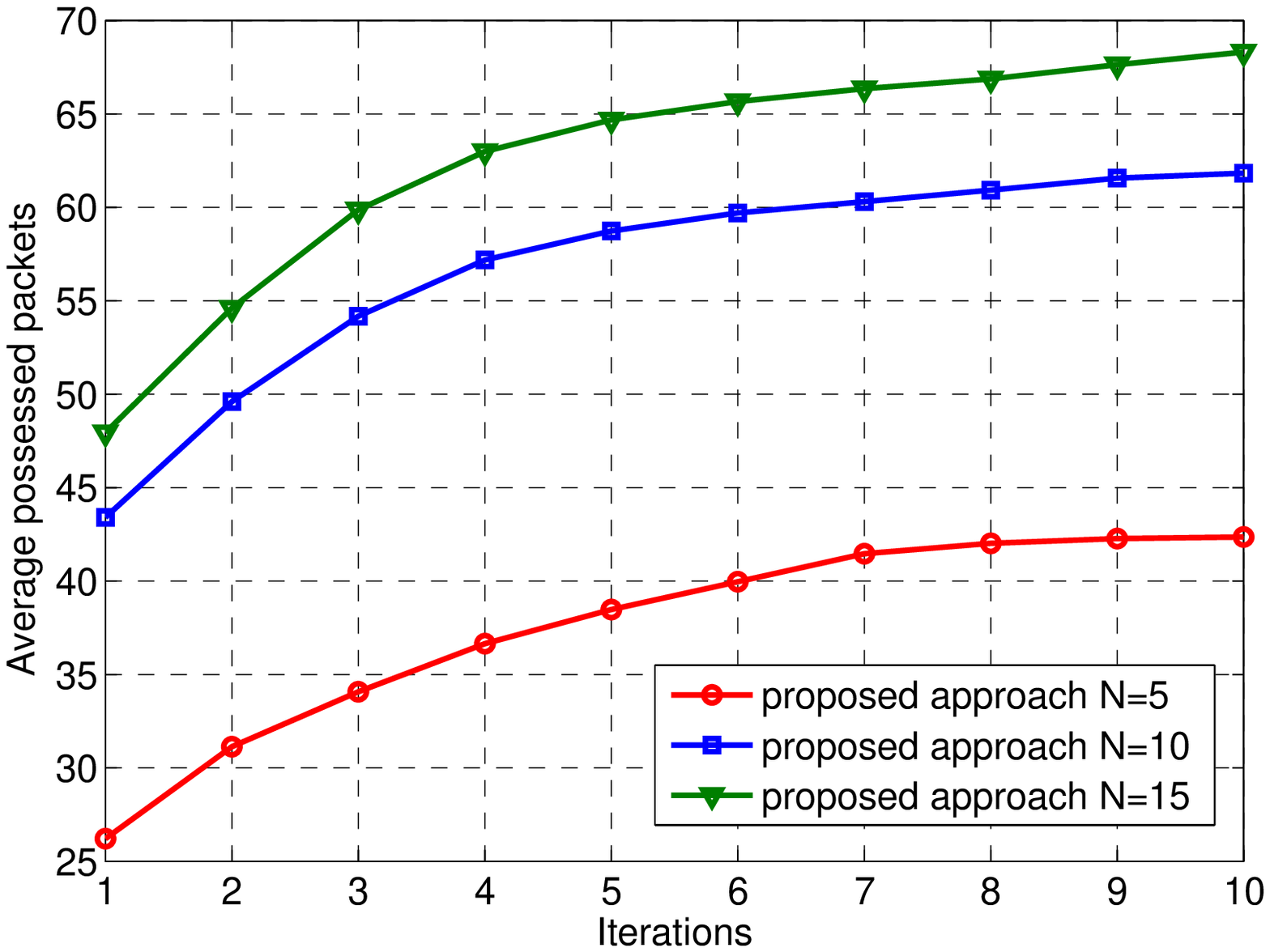}
\caption{Average possessed packets as a function of number of iterations by the proposed algorithm for networks with $N=5,10,15, K=10$ and $K'=4$.} \label{iteration}
\end{figure}

In Fig.~\ref{iteration}, we show the convergence performance of the proposed approach for networks with $N=5,10,15,K=10$ and $K'=4$. In Section \uppercase\expandafter{\romannumeral4}.B, we have proved the convergence of the proposed algorithm. As we can see, the proposed approach converges in a fast speed with different network sizes. Also, we can see the average possessed packets converge to different values with different network sizes. On the one hand, denser networks have more chance for V2V transmissions, and thus, brings more missing packets to the OBU after sufficient iterations of our proposed algorithm. On the other hand, since there is a minimal distance constraint between any two subsequent vehicles, the network can not be infinitely dense, and thus, the converging value of average possessed packets can not be infinitely large. Actually, we can see that the gap between $N=15$ and $N=10$ is considerably smaller than the gap between $N=5$ and $N=10$, which stands for a limited converging value when $N$ is infinite.

Also, if the iteration number is denoted by $I$, we have he overall complexity of the proposed algorithm is ${\rm O}(NI)$. Since the proposed algorithm is distributed and the overall complexity is shared by all OBUs, the calculation complexity of each OBU is ${\rm O}(I)$, which increases linearly with the iteration number. Thus, Fig.~6 also shows the performance of the proposed algorithm as a function of the calculation complexity of each OBU, where the performance converges rapidly to the maximal value as the individual calculation complexity increases.

\section{Conclusions}%

In this paper, we have addressed the PCD problem in VANETs, in which the RSU broadcasts a popular file to the passing OBUs, but the OBUs fail to receive some packets due to high speeds and channel fadings. To support reliable transmissions, we have proposed a P2P approach based on coalitional graph game to allow the OBUs to exchange data and complement the missing packets. Specifically, we have introduced a coalitional graph game to model the OBUs, and have proposed a coalition formation algorithm to implement the P2P approach. The convergence of the proposed algorithm has been proved and the overhead has be localized for arbitrary network sizes. Also, CR has been utilized to perform the P2P transmissions over unlicensed channels and the sensing-throughput tradeoff has been analyzed, in which the optimal number of sensed channels $K'$ satisfies $(K'+1) \ln{(K'+1)} + K' = K$, where $K$ is the number of potential channels. The simulation results show that, by introducing cooperation among OBUs in a coalitional graph game model, the maximal total throughput is increased by $133\%$ and $250\%$, relative to the non-cooperative approach and the pure broadcasting scheme. Also, for networks in a large scale, the total possessed packets of our proposed algorithm can be increased by $25\%$ and $218\%$, relative to the non-cooperative approach and the pure broadcasting scheme, respectively.



\begin{table}[!t]
\renewcommand{\arraystretch}{2.0}
\caption{Proposed Algorithm for Popular Content Distribution in Cognitive Radio VANETs} \label{NetworkFormation} \centering
\begin{tabular}{p{150mm}}

\hline

\textbf{Phase \uppercase\expandafter{\romannumeral1}: Spectrum Sensing}  \\

\quad Each OBU $i \in \mathcal{N}$ randomly senses $K_i$ unauthorized channels, the reference value of which is given by (\ref{KjDecide2}). \\

\textbf{Phase \uppercase\expandafter{\romannumeral2}: Network Formation }  \\

$\ast$ \textbf{repeat} \\
\quad In iteration $t+1$, given the current transmission graph $G^t(\mathcal{V},\mathcal{E}^t)$ (In the first iteration, $G^0$ is the final transmission graph of the last slot), a randomly chosen OBU $i \in \mathcal{N}$ engages in the algorithm as follows:
    \begin{enumerate}
    \item asks its ``neighbors" in $\mathcal{N}_i$ for the needed information for calculation.
    \item calculates the set of feasible strategies $F_i$ defined in definition \ref{def1}.
    \item updates the set of feasible strategies by ${F_i}'$ as in (\ref{UpdateFeasible}).
    \item chooses the local best response $s^*_i = (a^*_i,b^*_i)$ defined in definition \ref{def2}, and updates $h^{t+1}(s^*_i) = h^t(s^*_i) + 1, h^{t+1}(s_i) = h^t(s_i), \forall s_i \ne s^*_i$.
    \item the new graph $G^{t+1}(\mathcal{V},\mathcal{E}^{t+1})$ is updated by $\mathcal{E}^{t+1} = (\mathcal{E}^{t} \backslash \{e_{a_i,i}, e_{i,b_i} \} ) \cup \{e_{a^*_i,i}, e_{i,b^*_i}\}$, where $e_{a_i,i}, e_{i,b_i} \in \mathcal{E}^t$.
    \end{enumerate} \\

$\ast$ \textbf{until} converges to a final graph $G^T$ after $T$ iterations. \\

\textbf{Phase \uppercase\expandafter{\romannumeral3}: Data Transmission}  \\

\quad For any OBU $i \in \mathcal{N}$ with the set of available channels $\mathcal{K}^*_i$, OBU $i$
    \begin{enumerate}
    \item transmits to OBU $b$, if $e_{i,b} \in \mathcal{E}^T, b \ne i$, through its best available channel $k^* \in \mathcal{K}^*_i$, where $R^{k^*}_{i,b} = \max \nolimits_{k \in \mathcal{K}^*_i} R^k_{i,b}.$
    \item receives from $a$ (OBU $a$, or the RSU), if $e_{a,i} \in \mathcal{E}^T, a \ne i$, through $a$'s best available channel $k^* \in \mathcal{K}^*_a$, where $R^{k^*}_{a,i} = \max \nolimits_{k \in \mathcal{K}^*_a} R^k_{a,i}.$
    \end{enumerate} \\

\textbf{The algorithm is run repeatedly every slot for adapting to environmental changes.} \\

\hline

\end{tabular}
\end{table}

\begin{table}
\centering
\caption{Parameters for Simulation} \label{Parameters}
\begin{tabular}{|c|c|}
\hline $N = 1 \sim 50$ & number of OBUs in the network\\
\hline $L = 50m \times N$ & initial length of the fleet of vehicles\\
\hline $D = 350m$ & initial distance from the RSU\\
\hline $v_{min} = 10 m/s$ & the minimal speed \\
\hline $v_{max} = 30 m/s$ & the maximal speed \\
\hline $d_{min} = 50 m$ & the security distance \\
\hline $d_{max} = 100 m$ & the maximum distance \\
\hline $a = 1 \sim 5 m/s^2$ & the acceleration \\
\hline $p = 0.1$ & the probability of changing speed \\
\hline $M = 100$ & number of packets of the entire file\\
\hline $Ms = 100Mb$ & the size of the entire popular file\\
\hline $K = 1 \sim 10$ & number of V2V channels \\
\hline $K' = 1 \sim K $ & number of sensed V2V channels \\
\hline $W = 75MHz$ & the bandwidth of the V2R channel\\
\hline $W'= 10MHz$ & the bandwidth of V2V channels\\
\hline $\beta = 15dB$ & signal-to-noise rate for V2R transmission\\
\hline $\beta'= 10dB$ & signal-to-noise rate for V2V transmission\\
\hline $R_0 = 5Mb/s$ & the broadcasting rate of the RSU\\
\hline $\gamma_{out} = 0.5, \gamma_{in} = 1, \gamma_{cost} = 0.1$ & the pricing factors\\
\hline $P_m = 0.1 $ & the possibility of missing\\
\hline $P_f = 0.1 $ & the possibility of false alarm\\
\hline
\end{tabular}
\end{table}


\begin{thebibliography}{50}

\bibitem{OW-2009}
S.~Olariu and M.~A.~C.~Weigle, \emph{Vehicular Networks: From Theory to Practice,} London, UK: Chapman \& Hall/CRC, Computer and Information Sciences Series, Mar.~2009.

\bibitem{HL-2008}
H.~Hartenstein and K.~P.~Laberteaux, ``A Tutorial Survey On Vehicular Ad Hoc Networks," \emph{IEEE Communications Magazine}, vol.~46, no.~6, pp.~164-171, Jun.~2008.

\bibitem{LYL-2011}
M.~Li, Z.~Yang, and W.~Lou, ``Codeon: Cooperative Popular Content Distribution for Vehicular Networks using Symbol Level Network Coding," \emph{IEEE Journal on Selected Areas in Communications}, vol.~29, no.~1, pp.~223-235, Jan.~2011.

\bibitem{LCPSL-2005}
K.~Lua, J.~Crowcroft, M.~Pias, R.~Sharma, and S.~Lim, ``A Survey and Comparison of Peer-to-Peer Overlay Network Schemes," \emph{IEEE Communications Surveys $\&$ Tutorials}, vol.~7, no.~2, pp.~72-93, Second Quarter, 2005.

\bibitem{NDPGS-2005}
A.~Nandan, S.~Das, G.~Pau, M.~Gerla, and M.~Y.~Sanadidi, ``Co-operative Downloading in Vehicular Ad-hoc Wireless Networks," in \emph{The Second Annual Conference on Wireless On demand Network Systems and Services (WONS)}, pp.~32-41, St. Moritz, Switzerland, Jan.~2005.

\bibitem{LPYPG-2006}
U.~Lee, J.~S.~Park, J.~Yeh, G.~Pau, and M.~Gerla, ``Code Torrent: Content Distribution using Network Coding in VANET," in \emph{ACM 1st International Workshop on Decentralized
Resource Sharing in Mobile Computing and Networking (MobiShare)}, Los Angeles, CA, Sep. 2006.

\bibitem{ZZJ-2009}
J.~Zhang, Q.~Zhang, and W.~Jia, ``VC-MAC: A Cooperative MAC Protocol in Vehicular Networks," \emph{IEEE Transactions on Vehicular Technology}, vol.~58, no.~3, pp.~1561-1571, March.~2009.

\bibitem{Myerson-1977}
R.~Myerson, ``Graphs and Cooperation in Games," \emph{Math. Oper. Res.}, vol.~2, pp.~225-229, Jun.~1977.

\bibitem{SHDHB-2009}
W.~Saad, Z.~Han, M.~Debbah, A.~Hjorungnes, and T.~Basar, ``Coalitional Game Theory for Communication Networks: A Tutorial," \emph{IEEE Signal Processing Magazine}, vol.~26, no.~5, pp.~77-97, Sep.~2009.

\bibitem{SHDH-2008}
W.~Saad, Z.~Han, M.~Debbah, and A.~Hjoungnes, ``Network Formation Games for Distributed Uplink Tree Construction in IEEE 802.16j Networks," in \emph{Proceedings of IEEE Global Communication Conference}, New Orleans, LA, Dec.~2008.

\bibitem{AJM-2007}
E.~Arcaute, R.~Johari, and S.~Mannor, ``Network Formation: Bilateral Contracting and Myopic Dynamics," \emph{Lecture Notes Computer Science}, vol.~4858, pp.~191-207, Dec.~2007.

\bibitem{802}
IEEE P802.11p/D3.0, ``Draft Amendment to Standard for Information Technology-Telecommunications and Information Exchange between Systems-Local and Metropolitan Area Networks-Specific Requirements -- Part 11: Wireless LAN Medium Access Control (MAC) and Physical Layer (PHY) Specifications-Amendment 7: Wireless Access in Vehicular Environment," 2007.

\bibitem{MMKTPBZKC-2011}
C.~F.~Mecklenbrauker, A.~F.~Molisch, J.~Karedal, F.~Tufvesson, A.~Paier, L.~Bernado, T.~Zemen, O.~Klemp, and N.Czink, ``Vehicular Channel
Characterization and Its Implications for Wireless System Design and Performance," in \emph{Proc. the IEEE}, vol.~99, no.~7, pp.~1189-1212, Jul.~2011.

\bibitem{MPGW-2006}
A.~Mahajan, N.~Potnis, K.~Gopalan, and A.~I.~A.~Wang, ``Urban Mobility Models for Vanets," in \emph{Proceedings of the 2nd IEEE
International Workshop on Next Generation Wireless Networks}, Bangalore, India, Dec.~2006.

\bibitem{DKTT-2008}
J.~Derks, J.~Kuipers, M.~Tennekes, and F.~Thuijsman, ``Local Dynamics in Network Formation," in \emph{Proceedings of Third World Congress of The Game Theory Society}, Evanston, IL, Jul.~2008.

\bibitem{JW-2002}
M.~O.~Jackson and A.~Watts, ``The Evolution of Social and Economic Networks," \emph{Journal of Economic Theory}, vol.~106, pp.~265-295, Nov.~2002.

\bibitem{DW-2005}
G.~Demange and M.~Wooders, \emph{Group Formation in Economics: Networks, Clubs and Coalitions,} Cambridge, UK: Cambridge University Press, Mar.~2005.

\bibitem{BO-1999}
T.~Basar and G.~J.~Olsder, \emph{Dynamic Noncooperative Game Theory.} SIAM Series in Classics in Applied Mathematics, 1999.

\bibitem{SMG-2001}
C.~U.~Saraydar, N.~B.~Mandayam, and D.~J.~Goodman, ``Pricing and Power Control in A Multicell Wireless Data Network," \emph{IEEE Journal on Selected Areas in Communications}, vol.~19, no.~10, pp.~1883-1892, Oct.~2001.

\bibitem{NH-2007}
D.~Niyato and E.~Hossain, ``QoS-aware Bandwidth Allocation and Admission Control in IEEE 802.16 Broadband Wireless Access Networks: A
Noncooperative Game Theoretic Approach," \emph{Elsevier Computer Networks}, vol.~51, no.~11, pp.~3305-3321, Aug.~2007.

\bibitem{BAS-2003}
C.~Bouroghain, D.~Agrawal, and S.~Suri, ``A Game Theoretic Framework for Incentives in P2p Systems," in \emph{Proceedings of 3rd International Conference Peer-to-Peer Computing}, Linkoping, Sweden, Sep.~2003.

\bibitem{SHBDH-2011}
W.~Saad, Z.~Han, T.~Basar, M.~Debbah, and A.~Hjorungnes, ``Network Formation Games Among Relay Stations in Next Generation Wireless Networks," \emph{IEEE Transactions on Communications}, vol.~59, no.~9, pp.~2528-2542, Sep.~2011.

\bibitem{SHZHBP-2012}
W.~Saad, Z.~Han, R.~Zheng, A.~Hjorungnes, T.~Basar, and H.~V.~Poor, ``Coalitional Games in Partition Form for Joint Spectrum Sensing and Access in Cognitive Radio Networks," \emph{IEEE Journal of Selected Topics in Signal Processing}, vol.~6, no.~2, pp.~195-209, Apr.~2012.

\end{thebibliography}
\end{document}